\documentclass[aps,prl,reprint,superscriptaddress,longbibliography,floatfix]{revtex4-2}
\usepackage{amsmath,amssymb,bm,graphicx,booktabs}
\usepackage{mathptmx}
\SetSymbolFont{symbols}{normal}{OMS}{cmsy}{m}{n}
\DeclareMathAlphabet{\mathcal}{OMS}{cmsy}{m}{n}
\usepackage{fix-cm}
\usepackage[hidelinks]{hyperref}
\newcommand{\hc}{\mathrm{H.c.}}

\begin{document}
\makeatletter\immediate\write\@auxout{\string\citation{ArticleTitleControl}}\makeatother
\title{Spontaneous Chirality and Collective Chern Bands in a Dipole-Conserving Bose System}
\author{Yan-Guang Yue}
\affiliation{Department of Physics and State Key Laboratory of Surface Physics, Fudan University, Shanghai 200433, China}
\author{Jie Lou}
\affiliation{Department of Physics and State Key Laboratory of Surface Physics, Fudan University, Shanghai 200433, China}
\author{Yan Chen}
\email{yanchen99@fudan.edu.cn}
\affiliation{Department of Physics and State Key Laboratory of Surface Physics, Fudan University, Shanghai 200433, China}
\affiliation{Shanghai Branch, Hefei National Laboratory, Shanghai 201315, China}
\begin{abstract}
We demonstrate spontaneous chiral dipole order and an isolated collective Chern band in a time-reversal-symmetric microscopic bosonic lattice model with exact dipole-moment conservation. In a controlled large-occupation limit at zero temperature, frustrated dipole exchange and three-particle conversion stabilize two time-reversed chiral condensates. The two dipole Goldstone modes remain gapless, while a separate finite-frequency band carries Chern numbers $-1$ and $+1$ in the opposite domains. We trace this topology to complex mode couplings in the chiral background that gap finite-frequency Dirac crossings. Site-resolved density drives preserve dipole-moment conservation and reconstruct the band Chern number from band-resolved absorption spectra in the quadratic theory.
\end{abstract}
\maketitle
\clubpenalty=10000
\widowpenalty=10000

\textit{Introduction.---}Conservation of particle number and dipole moment fixes the center of mass but allows particle--hole dipoles to move~\cite{Pretko2017Subdimensional,Pretko2018Gauge,Gromov2019Multipole,Gromov2024Review}. The constraint can fragment Hilbert space~\cite{Sala2020,Khemani2020} and restrict excitation propagation~\cite{Oh2024,Boesl2024}. Experiments in tilted optical lattices have revealed constrained two-dimensional dynamics~\cite{Adler2024} and dipole quantum walks~\cite{Kim2026}. These dipoles can also form a condensate without single-particle condensation~\cite{Yuan2020,Chen2021FractonicII,Lake2022,Lake2023Tilted}. Their collective behavior includes dipole Luttinger liquids and dipole supersolids~\cite{Zechmann2023}, while multipolar condensates support Josephson effects without particle flow~\cite{Xu2024Multipolar}.

In ordinary Bose condensates, spontaneous chiral order can generate topological Bogoliubov bands~\cite{Furukawa2015,Bardyn2016,Xu2016,DiLiberto2016,Wang2021,Jalali2023}. Dipole constraints also connect to topology through insulating phases~\cite{MayMann2021,Lam2024} and fractional quantum Hall correlations in tilted bosonic chains~\cite{Sable2026}. For excitations, synthetic tensor gauge fields produce dipole Chern bands through externally imposed phases~\cite{Zhang2025Tensor}, while fermionic band topology can modify dipole Goldstone dynamics~\cite{Anakru2024}. Weak explicit breaking of dipole conservation provides another route, converting a dipole phase doublet into topological pseudo-Goldstone modes~\cite{Yue2026PseudoGoldstone}.

Can an exactly dipole-conserving, time-reversal-symmetric Bose system spontaneously develop chiral dipole order while retaining its gapless Goldstone modes and supporting a separate collective Chern band? Can this topology be reconstructed using probes that respect the same conservation law?

Here we construct a microscopic bosonic model realizing this structure in a controlled large-occupation limit. Frustrated dipole exchange and three-particle conversion stabilize two time-reversed chiral condensates. Unlike the pseudo-Goldstone topology induced by weak explicit breaking of dipole conservation~\cite{Yue2026PseudoGoldstone}, the two Goldstone modes remain gapless, while an isolated finite-frequency band carries a nonzero Chern number. We trace this topology to complex collective-mode couplings that gap finite-frequency Dirac crossings. Moreover, site-resolved density absorption reconstructs the integer band invariant without violating dipole-moment conservation.

\begin{figure}[t]
 \centering\includegraphics[width=\columnwidth]{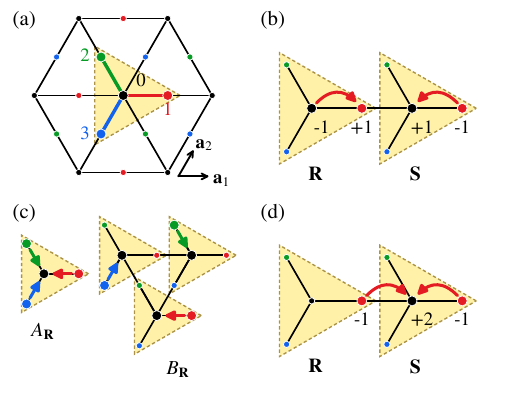}
 \caption{Microscopic lattice and particle processes. (a) Lattice and four-site basis. Black circles mark central sites; red, green, and blue mark outer sites $\nu=1,2,3$. Shaded regions group the four sites assigned to each cell, as labeled $0,1,2,3$ in (a). Black lines guide the eye. (b) Dipole exchange $k$: two particles move in opposite directions within their respective cells. (c) Three-particle conversion $j_A,j_B$: $A_{\mathbf R}$ and $B_{\mathbf R}$ move three particles inward, within one cell or across three cells. (d) Correlated pair hopping $t$: two particles move to their common midpoint. Panels (b,d) show the same neighboring cells $\mathbf R$ and $\mathbf S=\mathbf R+\mathbf a_1$, with $\nu=1$; signed numbers give occupation changes. Arrows within each process act simultaneously, and their colors identify the dipole orientation. Hermitian-conjugate processes reverse all arrows.}
 \label{fig:model}
\end{figure}

\textit{Microscopic model and conservation laws.---}We work throughout at zero temperature and set $\hbar=1$. Bosons occupy a triangular lattice with four sites per primitive cell [Fig.~\ref{fig:model}(a)]. Its primitive vectors are $\mathbf a_1=a(1,0)$ and $\mathbf a_2=a(1/2,\sqrt3/2)$. The reciprocal vectors satisfy $\mathbf a_i\cdot\mathbf b_j=2\pi\delta_{ij}$. We label momenta by reduced coordinates $\mathbf q=(q_1,q_2)$, with physical wave vector $\mathbf k=q_1\mathbf b_1+q_2\mathbf b_2$. Each cell $\mathbf R$ has a central site $\mathbf r_{\mathbf R0}=\mathbf R$ and three outer sites $\mathbf r_{\mathbf R\nu}=\mathbf R+\mathbf e_\nu/2$, with $\mathbf e_1=\mathbf a_1$, $\mathbf e_2=\mathbf a_2-\mathbf a_1$, and $\mathbf e_3=-\mathbf a_2$.
The operator $b_{\mathbf R\alpha}$ annihilates a boson on site $\alpha=0,1,2,3$, and $n_{\mathbf R\alpha}=b_{\mathbf R\alpha}^\dagger b_{\mathbf R\alpha}$. Define $N_{\mathbf R}=\sum_\alpha n_{\mathbf R\alpha}$ and the composite dipole $O_{\mathbf R\nu}=b_{\mathbf R0}^{\dagger}b_{\mathbf R\nu}$. The latter transfers one particle from outer site $\nu$ to the center, changing the particle dipole moment by $-\mathbf e_\nu/2$.

The Hamiltonian, with real coefficients and reference occupation $N$ per cell, is
\begin{align}
 H={}&H_d+H_C-\frac{k}{6N}\sum_{\mathbf R,\boldsymbol\delta,\nu}
 (O_{\mathbf R\nu}^{\dagger}O_{\mathbf R+\boldsymbol\delta,\nu}+\hc)
 \nonumber\\
 &-\frac1{N^2}\sum_{\mathbf R}(j_A A_{\mathbf R}+j_B B_{\mathbf R}+\hc)
 -\frac{t}{N}\sum_{\mathbf R,\nu}(T_{\mathbf R\nu}+\hc),
 \label{eq:H}\\
 H_d={}&\sum_{\mathbf R}\Bigl[\frac{u}{2N}\sum_\alpha n_{\mathbf R\alpha}(n_{\mathbf R\alpha}-1)
 +\delta_0(n_{\mathbf R0}-N_{\mathbf R}/2)
 \nonumber\\
 &\hspace{26mm}+\frac{v}{N}n_{\mathbf R0}\sum_\nu n_{\mathbf R\nu}\Bigr],
 \label{eq:Hd}\\
 H_C={}&N\sum_{\mathbf R}\Bigl[\frac{\Delta}{2}(N_{\mathbf R}-N)^2+w(N_{\mathbf R}-N)^6\Bigr].
 \label{eq:HCmain}
\end{align}
In $H_d$, $u$ is the on-site interaction, $\delta_0$ is the central-to-outer site energy offset, and $v$ couples the central and outer densities. The positive $\Delta$ and $w$ terms penalize changes in cell occupation; the sixth-power term stabilizes the energy at large occupation.

The $k$ term exchanges dipoles of the same orientation between neighboring cells, with $\boldsymbol\delta\in\{\mathbf a_1,\mathbf a_2,\mathbf a_1-\mathbf a_2\}$. In particle language, one boson moves outward in one cell while another moves inward in the other [Fig.~\ref{fig:model}(b)]. The $j_A$ and $j_B$ terms simultaneously transfer three bosons from outer to central sites, within one cell or across three cells [Fig.~\ref{fig:model}(c)]. They are $A_{\mathbf R}=\prod_\nu O_{\mathbf R\nu}$ and $B_{\mathbf R}=\prod_\nu O_{\mathbf R+\mathbf p_\nu,\nu}$, where $(\mathbf p_1,\mathbf p_2,\mathbf p_3)=(0,\mathbf a_2,\mathbf a_2-\mathbf a_1)$. Their Hermitian conjugates reverse these processes.

The $t$ term is correlated pair hopping [Fig.~\ref{fig:model}(d)]. With $\mathbf S=\mathbf R+\mathbf e_\nu$, the operator $T_{\mathbf R\nu}=(b_{\mathbf S0}^{\dagger})^2b_{\mathbf R\nu}b_{\mathbf S\nu}$ moves one particle from each of $(\mathbf R,\nu)$ and $(\mathbf S,\nu)$ to $(\mathbf S,0)$. It changes $(N_{\mathbf R},N_{\mathbf S})$ by $(-1,+1)$, allowing particles to move between cells.

Every process creates and annihilates the same number of particles, conserving $Q=\sum_{\mathbf R,\alpha}n_{\mathbf R\alpha}$. The dipole moment is $\mathbf P=\sum_{\mathbf R,\alpha}\mathbf r_{\mathbf R\alpha}n_{\mathbf R\alpha}$. The density terms $H_d$ and $H_C$ preserve it because they are diagonal in occupation. The $k$ process has opposite displacements $\pm\mathbf e_\nu/2$, and the three displacements in either $j$ process sum to $-\sum_\nu\mathbf e_\nu/2=0$. In the $t$ process, the two initial positions satisfy $\mathbf r_{\mathbf R\nu}+\mathbf r_{\mathbf S\nu}=2\mathbf r_{\mathbf S0}$. Consequently, $[H,Q]=[H,P_x]=[H,P_y]=0$ term by term. The real coefficients also preserve time reversal.

\textit{Spontaneous chiral dipole order.---}When changing a cell's occupation costs much more energy than redistributing particles within it, we eliminate these high-energy states while retaining the leading virtual $t$ correction. The resulting effective dipole theory describes density redistribution and relative phases within cells. Virtual changes in cell occupation generate a density interaction of strength $g=t^2/(\Delta+2w)$. We then use a large-occupation expansion to determine the ordered state and its quadratic fluctuations.

\begin{figure}[!htbp]
 \centering\includegraphics[width=\columnwidth]{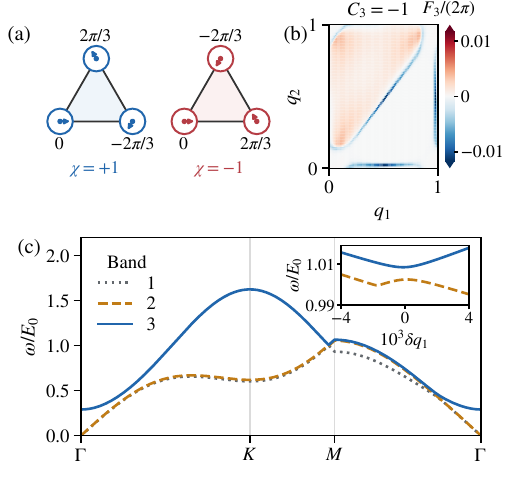}
 \caption{Spontaneous chiral order and a collective Chern band. (a) Phase patterns for $\chi=\pm1$. Each triangle shows one fixed dipole component at cells $\mathbf R$, $\mathbf R+\mathbf a_1$, and $\mathbf R+\mathbf a_2$ (lower left, lower right, top). Clock hands represent its complex phase, with the common offset set to zero. Time reversal exchanges the two patterns. (b,c) Benchmark parameters (SM), $g=0.1E_0$ and $\chi=+1$; $E_0$ is the energy unit. (b) Band-3 Berry flux per plaquette in units of $2\pi$, $F_3/(2\pi)$, on a $96\times96$ reciprocal mesh, with a symmetric logarithmic color scale. The plotted values sum to $C_3=-1$; the time-reversed domain has $C_3=+1$. (c) Positive-frequency bands along $\Gamma$--$K$--$M$--$\Gamma$. The two lower modes meet at zero frequency; $\omega_3(\Gamma)=0.290E_0$. Inset: bands 2 and 3 near their minimum direct separation $\Delta_{23}=6.06\times10^{-3}E_0$, along $\mathbf q=\mathbf q_{\min}+(\delta q_1,0)$, with $\mathbf q_{\min}\simeq(0.514,0.490)$.}
 \label{fig:topology}
\end{figure}

The dipole condensate is described by $D_{\mathbf R\nu}\equiv\langle O_{\mathbf R\nu}\rangle/N=|D_{\mathbf R\nu}|e^{i\phi_{\mathbf R\nu}}$, where $\phi_{\mathbf R\nu}$ is the phase of the coherence between the central site and outer site $\nu$. The $k$ term couples these phases between neighboring cells for each dipole orientation. At fixed magnitudes, the bond energy varies as $E_{k,\mathrm{bond}}\propto-k\cos(\phi_{\mathbf R'\nu}-\phi_{\mathbf R\nu})$. For $k<0$, each bond favors a phase difference of $\pi$, which cannot be satisfied on all three sides of a triangle. At equal magnitudes, the minimum instead has phases separated by $120^\circ$ [Fig.~\ref{fig:topology}(a)]. These are phases of the same dipole component in neighboring cells. The two opposite arrangements have ordering wavevectors with reciprocal coordinates $\pm\mathbf Q$, where $\mathbf Q=(2/3,1/3)$. For $+\mathbf Q$, translation by $\mathbf a_1$ or $\mathbf a_2$ changes the phase by $4\pi/3$ or $2\pi/3$, respectively; $-\mathbf Q$ reverses these changes.

The three-particle conversion coordinates the choices of the three dipole components. At fixed magnitudes, the intracell term contributes $E_{A,\mathbf R}\propto-j_A\cos(\phi_{\mathbf R1}+\phi_{\mathbf R2}+\phi_{\mathbf R3})$. Thus $j_A>0$ locks the sum of the three component phases within each cell to zero modulo $2\pi$. Choosing $+\mathbf Q$ for all three components, or $-\mathbf Q$ for all three, satisfies this lock throughout the lattice because $3\mathbf Q$ has integer reciprocal coordinates. Mixing the two choices leaves a spatially varying phase sum. The intercell term $j_B>0$ is also minimized for either common choice. Writing $\mathbf R=n_1\mathbf a_1+n_2\mathbf a_2$, the resulting condensate is
\begin{equation}
 D_{\mathbf R\nu}=D_0e^{2\pi i\chi(2n_1+n_2)/3+i\theta_\nu},
 \qquad \sum_\nu\theta_\nu=0,
 \label{eq:order}
\end{equation}
where $D_0>0$ is the common magnitude, $\chi=\pm1$ labels the two winding senses, and $\theta_\nu$ are position-independent phase offsets.

Time reversal complex-conjugates $D_{\mathbf R\nu}$ and exchanges the two winding senses. The real Hamiltonian leaves them degenerate, so selecting one spontaneously breaks time reversal.

The two free phase offsets follow from the exact dipole symmetries. A rotation $e^{i\boldsymbol\beta\cdot\mathbf P}$, with two real parameters $\boldsymbol\beta$, sends $O_{\mathbf R\nu}\to e^{-i\boldsymbol\beta\cdot\mathbf e_\nu/2}O_{\mathbf R\nu}$ and preserves the locked phase sum because $\sum_\nu\mathbf e_\nu=0$. The condensate breaks both dipole symmetries while particle-number symmetry remains intact~\cite{Stahl2022Multipole}. The commuting generators $P_x,P_y$ therefore give two Goldstone modes~\cite{WM2012}.

In the large-occupation limit, this effective dipole theory establishes the chiral state as the minimum of the leading coherent-state energy for the parameters used here. Small fluctuations are stable at every momentum. The derivation and validity conditions are given in the Supplemental Material (SM~\cite{SM}).

\textit{Topological collective excitations.---}The chiral condensate supports two gapless Goldstone modes and an isolated gapped collective band. Factoring out the ordered phase in Eq.~\eqref{eq:order} leaves a quadratic problem periodic in the original cell. Within this effective theory, the four-site cell has three independent relative phases and their conjugate density fluctuations. Their coupled oscillations give three collective modes. Here $\mathbf q$ labels the spatial modulation of a fluctuation about the ordered pattern. Uniform shifts of the two free phase combinations give the Goldstone modes at $\mathbf q=0$ ($\Gamma$). The locked phase sum supplies a restoring force for the remaining combination.

We obtain the mode frequencies and eigenvectors from the Bogoliubov--de Gennes (BdG) equations, which describe quadratic (Gaussian) fluctuations~\cite{Colpa1978,Chaudhary2021}. For the $g=0.1E_0$ benchmark, the third band has a nonzero direct gap from the lower bands at every momentum [Fig.~\ref{fig:topology}(c)]. Integrating its bosonic Berry curvature gives $C_3=(2\pi)^{-1}\int_{\mathrm{BZ}}\Omega_3(\mathbf q)\,d^2q=-1$ in the $\chi=+1$ domain~\cite{Shindou2013,FHS2005} [Fig.~\ref{fig:topology}(b)]. Time reversal maps $(\chi,\mathbf q)$ to $(-\chi,-\mathbf q)$ and reverses the curvature, giving $C_3=+1$ in the opposite domain.

\begin{figure}[!htbp]
 \centering\includegraphics[width=\columnwidth]{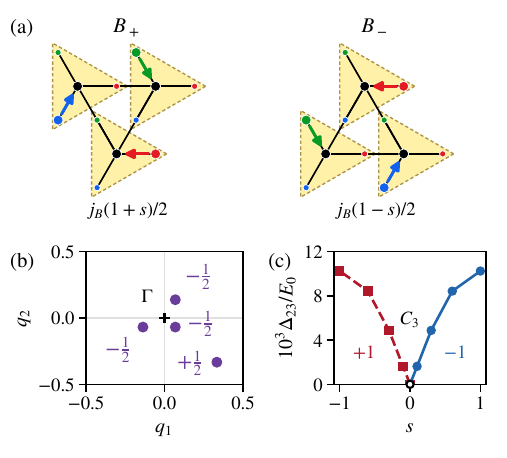}
 \caption{Origin of the collective Chern band. Mechanism reference, $g=0$, $\chi=+1$. (a) Intercell conversion terms $B_\pm=\prod_\nu O_{\mathbf R\pm\mathbf p_\nu,\nu}$ have weights $j_B(1\pm s)/2$. Reversing the cell offsets preserves the physical orientation $\mathbf e_\nu/2$ of each dipole. Lattice lines, shaded four-site groups, and simultaneous-transfer arrows follow Fig.~1. (b) Four band-2/3 contacts at $s=0$, displayed modulo reciprocal lattice vectors with $q_i\in[-1/2,1/2)$. Labels give their upper-band local flux contributions, in units of $2\pi$, for small positive $s$. (c) Minimum direct gap at fixed chirality. Displayed points have $C_3=+1$ for $s<0$ and $-1$ for $s>0$; at $s=0$ the gap closes and the single-band invariant is undefined. Negative-$s$ points follow from complex conjugation; lines guide the eye. The ordered-state energy and exact conservation law are unchanged.}
 \label{fig:mechanism}
\end{figure}

Expanding the real microscopic couplings about the chiral condensate produces complex couplings between collective modes. To isolate how these generate topology, we use a reference without the virtual-hopping correction, $g=0$, and keep $\chi=+1$ fixed. We vary the intercell conversion by assigning weights $j_B(1+s)/2$ and $j_B(1-s)/2$ to the patterns with offsets $\mathbf p_\nu$ and $-\mathbf p_\nu$ [Fig.~\ref{fig:mechanism}(a)]. At $s=0$ the weights are equal; at $s=1$ the original $B$ process is recovered. This interpolation changes the collective-mode coupling while preserving microscopic time reversal, dipole-moment conservation, and the chiral state's energy.

At equal weights, bands 2 and 3 meet at four finite-frequency Dirac points [Fig.~\ref{fig:mechanism}(b)]. Near each point, their coupling is described to first order in $\delta\mathbf q$ and $s$ by
\begin{equation}
 \begin{split}
 h_D={}&[\omega_D+\mathbf v_0\cdot\delta\mathbf q]I\\
 &+(\mathbf v_1\delta q_1+\mathbf v_2\delta q_2+s\mathbf m)\cdot\boldsymbol\sigma,
 \end{split}
 \label{eq:Dirac}
\end{equation}
Here $\delta\mathbf q$ is measured from the crossing, $\omega_D$ is its frequency, and $\mathbf v_0$ sets the common spectral tilt. The Pauli matrices $\boldsymbol\sigma$ act on the two modes; $\mathbf v_{1,2}$ and $\mathbf m$ are three-component coefficient vectors in this Pauli basis. In a real two-mode basis, $\mathbf v_{1,2}$ describe the momentum-dependent frequency difference and real mixing. The conversion imbalance controls the imaginary intermode coupling $s\mathbf m\cdot\boldsymbol\sigma$, which acts as a Dirac mass and gaps the crossings. Together with the momentum-dependent mixing, it produces Berry curvature.

For small positive $s$, one gapped Dirac point contributes $+1/2$ and the other three each contribute $-1/2$ to the upper-band Berry flux in units of $2\pi$. These local contributions account for the integer obtained from the full Brillouin zone. Within this reference, complex conjugation under $s\to-s$ reverses the Chern number at fixed condensate chirality [Fig.~\ref{fig:mechanism}(c)].

At $s=1$, we restore the virtual-hopping correction from $g=0$ to $g=0.1E_0$. Reoptimizing the condensate at each sampled value retains a nonzero direct gap and the Chern number, supporting a gap-preserving connection to the benchmark.

At $g=0$, changing $v$ and $\delta_0$ gives a chiral control with $C_3=0$ (SM). Chiral order allows complex mode couplings; their detailed structure determines whether the band is topological.

\textit{Conserving density spectroscopy.---}We probe the collective Chern band through the density motion within each cell. A uniform force couples to the conserved dipole moment and cannot induce finite-frequency absorption. Smooth total-density probes are also suppressed at long wavelengths~\cite{Zechmann2024}. We therefore modulate the three outer-site potentials independently with $H_f=\epsilon e^{-i\omega t}\sum_\nu f_\nu n_\nu(-\mathbf q)+\hc$. Here $\epsilon$ is the weak drive amplitude, $\omega$ its frequency, and $n_\nu(\mathbf q)$ the cell-periodic Fourier density. The complex coefficients $f_\nu$ set the modulation amplitudes and temporal phases. These density couplings conserve particle number and dipole moment, providing a complementary route to tensor-electric driving~\cite{Zhang2026Resonant}.

\begin{figure}[!htbp]
 \centering\includegraphics[width=\columnwidth]{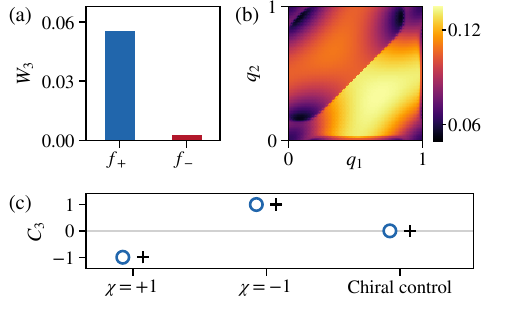}
 \caption{Density readout of collective topology. (a) Absolute band-3 peak areas for the drive patterns $f_\pm=(1,\pm i,0)/\sqrt2$ at $\mathbf q=(0.840,0.819)$ in the $\chi=+1$ domain. Time reversal exchanges the patterns at opposite momentum. (b) Summed site-resolved weight $\operatorname{Tr}S_3=\sum_\nu|d_{3\nu}|^2$, nonzero throughout the Brillouin zone. (c) Nine-drive reconstruction (circles) and BdG calculation (crosses) for the two benchmark domains and the chiral trivial control; markers are offset horizontally for visibility. All panels use noiseless, band-resolved weights from the quadratic theory; no linewidth is imposed. The two-pattern contrast in (a) alone does not determine topology.}
 \label{fig:density}
\end{figure}

At fixed $\mathbf q$, scanning the drive frequency reveals the collective modes in the absorption spectrum. For a resolved Bogoliubov mode, the peak area under a common drive calibration is $W_n(f)=f^\dagger S_n f$, where $S_n=d_nd_n^\dagger$ and $d_n$ collects its three site-density transition amplitudes. Opposite relative phases give different peak areas [Fig.~\ref{fig:density}(a)]. Three single-site drives and six pairwise drives with relative phases $0$ or $\pi/2$ determine $S_n$. This procedure uses interference between density drives, following spectroscopic and tomographic approaches to band geometry~\cite{Flaschner2016,Tran2017,Ozawa2018,Asteria2019,Tesfaye2025}.

For nonzero $d_n$, the reconstructed matrix $S_n=d_nd_n^\dagger$ has rank one, so $P_n=S_n/\operatorname{Tr}S_n$ projects onto its density-response vector. In our coherent-state description, all four sites have nonzero equilibrium occupation fractions. Together with positive Bogoliubov norm, this ensures a nonzero density response at every momentum on the isolated band [Fig.~\ref{fig:density}(b)]. Under a phase change of the mode, its density-response vector acquires the same phase. This smooth, nonvanishing projection therefore preserves the Chern number~\cite{Huang2018},
\begin{equation}
 C_n^{\rm density}=C_n.
 \label{eq:tomography}
\end{equation}
Its local curvature can differ from the Bogoliubov Berry curvature, but the integral gives the same integer. Applied across the Brillouin zone, the nine-drive reconstruction yields $-1$ and $+1$ for the two chiral domains and $0$ for the chiral trivial control [Fig.~\ref{fig:density}(c)]. The reconstruction uses peak areas and known drive patterns.

The proof, reconstruction formulas, and finite-occupation self-energy analysis~\cite{Beliaev2024} are given in SM. Beyond quadratic theory, it remains unresolved whether finite-$N$ corrections preserve a distinct third-band quasiparticle peak at every momentum.

\textit{Discussion.---}Exact dipole-moment conservation fixes the microscopic particle center of mass while allowing nontrivial collective-band dynamics. In our model, spontaneous chiral order supports an isolated Chern band alongside two Goldstone modes, extending the physics of dipole condensates beyond their long-wavelength hydrodynamics~\cite{Jain2023,Stahl2023Hydro,Armas2024Ideal}. Conservation also determines which probes can reveal this topology: uniform forces produce no finite-frequency absorption, whereas relative site potentials resolve internal density motion that retains the band's Chern number. Dipole conservation thus constrains both the collective excitations and the observables through which their topology can be measured.

\raggedbottom
\textit{Acknowledgments.---}We thank Fei Teng for his helpful discussions. This work is supported by the National Key Research and Development Program of China (Grant No. 2022YFA1404204), the National Natural Science Foundation of China (Grant No. 12274086), and the Quantum Science and Technology-National Science and Technology Major Project (Grant No. 2024ZD0300104).

\bibliographystyle{apsrev4-2}
\bibliography{references}
\end{document}


\makeatletter\immediate\write\@auxout{\string\citation{ArticleTitleControl}}\makeatother
\maketitle

This Supplemental Material specifies the microscopic particle model, the controlled effective dipole theory, the energy bound, and the collective-band calculations. It then derives density-spectrum tomography and the conserving energy response, and quantifies finite-occupation and finite-charge-gap corrections. All calculations are at zero temperature, with $\hbar=1$ unless restored explicitly. The benchmark is $g=0.1$ unless stated otherwise; the mechanism reference with mirrored conversion patterns uses $g=0$.

\section{Microscopic model and projection onto fixed cell occupations}
\subsection{Physical sites and interactions}
Choose $\mathbf a_1=a(1,0)$ and $\mathbf a_2=a(1/2,\sqrt3/2)$, with reciprocal basis $\mathbf a_i\cdot\mathbf b_j=2\pi\delta_{ij}$. The reduced momentum $\mathbf q=(q_1,q_2)$ labels the physical wave vector $\mathbf k=q_1\mathbf b_1+q_2\mathbf b_2$. For $\mathbf R=n_1\mathbf a_1+n_2\mathbf a_2$, the physical phase is $\mathbf k\cdot\mathbf R=2\pi(q_1n_1+q_2n_2)$. Below, lattice vectors are also written as integer coordinate pairs; $\mathbf q\cdot\mathbf R$ denotes the coordinate pairing $q_1n_1+q_2n_2$, so Fourier phases are $e^{\pm2\pi i\mathbf q\cdot\mathbf R}$. The three directed cell translations and the intercell conversion offsets are
\begin{equation}
 \mathbf e_1=(1,0),\quad\mathbf e_2=(-1,1),\quad\mathbf e_3=(0,-1),
 \qquad \mathbf p_1=(0,0),\quad\mathbf p_2=(0,1),\quad\mathbf p_3=(-1,1).
\end{equation}
Physical sites are $\mathbf r_{\mathbf R0}=\mathbf R$ and $\mathbf r_{\mathbf R\nu}=\mathbf R+\mathbf e_\nu/2$. They are distinct sites; the construction has four orbitals per triangular cell. The dipoles $O_{\mathbf R\nu}=b_{\mathbf R0}^{\dagger}b_{\mathbf R\nu}$ are physical bilinears, not independent canonical bosons.

The full microscopic Hamiltonian is
\begin{align}
 H_n={}&\sum_{\mathbf R}\left[\frac{u}{2N}\sum_{\alpha=0}^3 n_{\mathbf R\alpha}(n_{\mathbf R\alpha}-1)
 +\frac{\delta_0}{2}\left(n_{\mathbf R0}-\sum_{\nu=1}^3n_{\mathbf R\nu}\right)
 +\frac{v}{N}n_{\mathbf R0}\sum_\nu n_{\mathbf R\nu}\right]\nonumber\\
 &-\frac{k}{6N}\sum_{\mathbf R,\boldsymbol\delta\in\{(1,0),(0,1),(1,-1)\},\nu}
 (O_{\mathbf R\nu}^{\dagger}O_{\mathbf R+\boldsymbol\delta,\nu}+\hc)\nonumber\\
 &-\frac{j_A}{N^2}\sum_{\mathbf R}(O_{\mathbf R1}O_{\mathbf R2}O_{\mathbf R3}+\hc)
 -\frac{j_B}{N^2}\sum_{\mathbf R}(O_{\mathbf R+\mathbf p_1,1}O_{\mathbf R+\mathbf p_2,2}O_{\mathbf R+\mathbf p_3,3}+\hc),
 \label{eq:Hn}\\
 H_C={}&N\sum_{\mathbf R}\left[\frac{\Delta}{2}(N_{\mathbf R}-N)^2+w(N_{\mathbf R}-N)^6\right],
 \qquad H_t=-\frac{t}{N}\sum_{\mathbf R\nu}(T_{\mathbf R\nu}+\hc),\nonumber\\
 T_{\mathbf R\nu}={}&(b_{\mathbf S0}^{\dagger})^2b_{\mathbf R\nu}b_{\mathbf S\nu},
 \qquad\mathbf S=\mathbf R+\mathbf e_\nu,\qquad H=H_n+H_C+H_t.
 \label{eq:HC}
\end{align}
The real parameters in units of $E_0$ are
\begin{equation}
 (u,k,j_A,j_B,v)=(4,-2.24,0.24,0.08,3.5),\qquad
 \delta_0=-\frac{457}{375}+\frac{8\sqrt7}{625},\qquad g=\frac{t^2}{\Lambda}=0.1,
 \quad\Lambda=\Delta+2w.
 \label{eq:params}
\end{equation}
Here charge denotes particle number; a cell-charge defect has $N_{\mathbf R}\ne N$. The sixth-power cell-occupation term bounds high local occupations and fixes the costs of multiple charge defects. Its value matters for the fourth-order projection below.

\subsection{Exact spatial conservation and removal of cell charges}
With $Q=\sum_{\mathbf R\alpha}n_{\mathbf R\alpha}$ and $\mathbf P=\sum_{\mathbf R\alpha}\mathbf r_{\mathbf R\alpha}n_{\mathbf R\alpha}$,
\begin{equation}
 [\mathbf P,O_{\mathbf R\nu}]=-\frac{\mathbf e_\nu}{2}O_{\mathbf R\nu},\qquad
 [Q,H]=[P_i,H]=0.
\end{equation}
In $T_{\mathbf R\nu}$ the created position sum is $2\mathbf S$, equal to $(\mathbf R+\mathbf e_\nu/2)+(\mathbf S+\mathbf e_\nu/2)$. Each term conserves the spatial dipole moment. Time reversal is complex conjugation in the physical occupation basis. Conservation is defined on an open sample or the infinite lattice; periodic cells are used to compute translation-invariant bulk spectra.

The $T$ term changes $N_{\mathbf R}$ by $-1$ and $N_{\mathbf S}$ by $+1$. More specifically, consider any continuous diagonal one-body charge $F=\sum f_{\mathbf R\alpha}n_{\mathbf R\alpha}$. Exchange conservation requires $f_{\mathbf R0}-f_{\mathbf R\nu}=c_\nu$, independent of $\mathbf R$, and three-particle conversion requires $\sum_\nu c_\nu=0$. Conservation under $T$ further gives
\begin{equation}
 f_{\mathbf R+\mathbf e_\nu,0}-f_{\mathbf R0}=-2c_\nu.
\end{equation}
Thus $f_{\mathbf R0}=A+\mathbf B\cdot\mathbf R$ and $f_{\mathbf R\nu}=A+\mathbf B\cdot(\mathbf R+\mathbf e_\nu/2)$: only $Q$ and the two components of $\mathbf P$ remain within this class. This statement does not classify discrete or nonlocal conserved operators. It establishes that the dipole order is not protected by an imposed set of independent cell-number symmetries in the full Hamiltonian.

\subsection{Second-order charge projection}
A single $T$ creates a pair of cell-charge defects whose leading energy is $N\Lambda$. Projection onto $N_{\mathbf R}=N$ gives
\begin{equation}
 H_{\mathrm{eff}}^{(2)}=-\frac{t^2}{N^3\Lambda}\sum_{\mathbf R\nu}
 (T_{\mathbf R\nu}^{\dagger}T_{\mathbf R\nu}+T_{\mathbf R\nu}T_{\mathbf R\nu}^{\dagger}).
 \label{eq:proj2}
\end{equation}
For one directed link,
\begin{align}
 T^{\dagger}T&=n_{\mathbf R\nu}n_{\mathbf S\nu}(n_{\mathbf S0}+1)(n_{\mathbf S0}+2),\nonumber\\
 TT^{\dagger}&=(n_{\mathbf R\nu}+1)(n_{\mathbf S\nu}+1)n_{\mathbf S0}(n_{\mathbf S0}-1).
\end{align}
Writing the occupation fractions as $p_{\mathbf R\alpha}=\langle n_{\mathbf R\alpha}\rangle/N$, their leading coherent-state energy divided by $N$ is
\begin{equation}
 E_g/N=-2g\sum_{\mathbf R\nu}p_{\mathbf R\nu}p_{\mathbf S\nu}p_{\mathbf S0}^2.
 \label{eq:Eg}
\end{equation}
At fixed $g$, increasing $\Lambda$ gives $t/\Lambda=\sqrt{g/\Lambda}\to0$. The simultaneous use of physical large occupation and charge projection is quantified in Sec.~\ref{sec:quantum}; $N$ is the physical occupation per cell~\cite{Jensen2024}.

\section{Global coherent-state minimum and spontaneous chirality}
\subsection{Leading energy and stationary state}
Use a fixed-particle-number coherent state in each cell,
\begin{equation}
 |z_{\mathbf R};N\rangle=\frac{(\sum_\alpha z_{\mathbf R\alpha}b_{\mathbf R\alpha}^{\dagger})^N}{\sqrt{N!}}|0\rangle,
 \qquad \sum_\alpha|z_{\mathbf R\alpha}|^2=1.
\end{equation}
Then $p_{\mathbf R\alpha}=|z_{\mathbf R\alpha}|^2$. Set $x_{\mathbf R}=|z_{\mathbf R0}|^2$, $y_{\mathbf R\nu}=|z_{\mathbf R\nu}|^2$, and $D_{\mathbf R\nu}=z_{\mathbf R0}^*z_{\mathbf R\nu}$. Let $L$ be the number of cells. The leading total energy divided by $N$, denoted $E_{\rm cl}$, is
\begin{align}
 E_{\rm cl}={}&\sum_{\mathbf R}\left[\frac u2(x_{\mathbf R}^2+\sum_\nu y_{\mathbf R\nu}^2)
 +\delta_0(x_{\mathbf R}-1/2)+vx_{\mathbf R}(1-x_{\mathbf R})\right]
 -\frac{k}{3}\sum_{\mathbf R\boldsymbol\delta\nu}\Rea(D_{\mathbf R\nu}^*D_{\mathbf R+\boldsymbol\delta,\nu})\nonumber\\
 &-2j_A\sum_{\mathbf R}\Rea\prod_\nu D_{\mathbf R\nu}
 -2j_B\sum_{\mathbf R}\Rea\prod_\nu D_{\mathbf R+\mathbf p_\nu,\nu}
 -2g\sum_{\mathbf R\nu}y_{\mathbf R\nu}y_{\mathbf S\nu}x_{\mathbf S}^2.
 \label{eq:Ecl}
\end{align}
The triangular structure factor
$\gamma(\mathbf q)=[\cos(2\pi q_1)+\cos(2\pi q_2)+\cos2\pi(q_1-q_2)]/3$
has minimum $-1/2$ at $\pm\mathbf Q$, $\mathbf Q=(2/3,1/3)$. For the common-valley state,
\begin{equation}
 z_{\mathbf R0}=\sqrt{x},\qquad z_{\mathbf R\nu}=\sqrt{y}\,e^{2\pi i\chi\mathbf Q\cdot\mathbf R+i\theta_\nu},
 \qquad y=(1-x)/3,\quad \sum_\nu\theta_\nu=0.
 \label{eq:z}
\end{equation}
Both conversion terms are phase compatible. Define $K_e=k\gamma(\mathbf Q)=|k|/2$ and $j=j_A+j_B$. Its energy per cell per reference particle, $e_g=E_{\rm cl}/L$, is
\begin{equation}
 e_g(x)=\frac u2(x^2+3y^2)+\delta_0(x-1/2)+(v-K_e)3xy-2j(xy)^{3/2}-6gx^2y^2.
 \label{eq:eg}
\end{equation}
For Eq.~\eqref{eq:params}, the minimum is at $x_*\simeq0.315$, with $e_*\simeq1.229E_0$. At $g=0$, the same microscopic detuning gives $x=0.3$.

For a positively oriented triangle $(\mathbf R,\mathbf R+\mathbf a_1,\mathbf R+\mathbf a_2)$, an order parameter invariant under dipole-symmetry phase transformations is
\begin{equation}
 \kappa_\nu=-\frac{2}{3\sqrt3}\Ima\left(
 D_{\mathbf R\nu}^*D_{\mathbf R+\mathbf a_1,\nu}
 +D_{\mathbf R+\mathbf a_1,\nu}^*D_{\mathbf R+\mathbf a_2,\nu}
 +D_{\mathbf R+\mathbf a_2,\nu}^*D_{\mathbf R\nu}\right)=\chi xy.
\end{equation}
It changes sign under time reversal. The two independent $\theta_\nu$ are phases associated with the two spontaneously broken dipole symmetries. Since the condensate operator carries no net particle number, particle-number symmetry is unbroken~\cite{Lake2022,Stahl2022Multipole}. The zero-temperature setting is essential in two dimensions~\cite{Kapustin2022}.

\subsection{Bound for arbitrary coherent configurations}
The exchange structure-factor bound gives $E_{\rm ex}/N\ge-K_e\sum_{\mathbf R}x_{\mathbf R}(1-x_{\mathbf R})$. For either conversion term,
\begin{equation}
 \Rea(D_1D_2D_3)\le\frac13\sum_\nu|D_\nu|^3.
\end{equation}
After relabeling the intercell sum, its bound is local. For the charge correction choose $r=x_*/y_*$ and apply the weighted arithmetic--geometric mean inequality
\begin{equation}
 ABC^2\le x_*^2y_*^2\left[\frac14(A/y_*)^4+\frac14(B/y_*)^4+\frac12(C/x_*)^4\right].
\end{equation}
Summing links produces $E_{\rm cl}\ge\sum_{\mathbf R}h(x_{\mathbf R},y_{\mathbf R\nu})$, where
\begin{align}
 h={}&\frac u2(x^2+\sum y_\nu^2)+\delta_0(x-1/2)+(v-K_e)x(1-x)
 -\frac{2j}{3}x^{3/2}\sum y_\nu^{3/2}\nonumber\\
 &-gr^2\sum y_\nu^4-3gr^{-2}x^4.
\end{align}
Let $s_x=1-x$ and $m=s_x/3$. On $\sum y_\nu=s_x$, the inequalities
\begin{equation}
 \sum y_\nu^{3/2}-3m^{3/2}\le\frac{\sum(y_\nu-m)^2}{2\sqrt m},\qquad
 \sum y_\nu^4-3m^4\le2s_x^2\sum(y_\nu-m)^2
\end{equation}
imply
\begin{align}
 h-B(x)&\ge A(x)\sum_\nu(y_\nu-m)^2,\nonumber\\
 A(x)&=\frac u2-\frac j3\frac{x^{3/2}}{\sqrt{(1-x)/3}}-2gr^2(1-x)^2,\nonumber\\
 B(x)&=e_0(x)-\frac{gr^2}{27}(1-x)^4-3gr^{-2}x^4.
 \label{eq:bound}
\end{align}
Here $e_0$ is Eq.~\eqref{eq:eg} at $g=0$. Conservative bounds give $A\ge1.120$ on $[0,0.9]$ and $A\ge0.176$ on $[0.9,0.99]$. For $x\ge0.99$, use $\sum y_\nu^{3/2}\le(1-x)^{3/2}$ and $\sum y_\nu^4\le(1-x)^4$ directly; they give $h\ge1.245E_0>e_*$.

The remaining scalar derivative is
\begin{align}
 B'(x)&=P_3(x)-\frac j{\sqrt3}(1-2x)\sqrt{x(1-x)},\nonumber\\
 P_3(x)&=\frac u3(4x-1)+\delta_0+(v-K_e)(1-2x)
 +\frac{4gr^2}{27}(1-x)^3-12gr^{-2}x^3.
\end{align}
Root isolation of $P_3^2-(j^2/3)(1-2x)^2x(1-x)$, followed by substitution into the unsquared derivative, gives two stationary points: $x_*\simeq0.315$ (minimum) and $x\simeq0.860$ (maximum). The other two roots in $[0,1]$ are introduced by squaring. The values at the endpoints and stationary points are $B(0)\simeq1.252E_0$, $B(x_*)\simeq1.229E_0$, $B(0.860)\simeq1.261E_0$, and $B(0.99)\simeq1.252E_0$, establishing the scalar minimum. This step uses 90-digit evaluation and Sturm isolation of a 70-digit rational approximation; it is a numerical certification of the scalar minimization, not an interval proof for exact algebraic coefficients.

All inequalities are saturated by Eq.~\eqref{eq:z}. Saturation fixes equal outer occupations, and confines each dipole Fourier component to $\pm\mathbf Q$. Constant amplitude excludes a superposition of both valleys within a dipole component. The $A$ conversion term then forces the three components to select the same valley; the $B$ conversion term is compatible with both choices. Hence the leading coherent ground states are the two chiral families with continuous dipole phases. This all-configuration bound implies a nonnegative coherent-state Hessian for every momentum. It establishes global stability at the leading large-$N$, second-order charge-projection level, not an exact finite-$N$ many-body ground-state theorem.

\begin{figure}[!htbp]
\centering\includegraphics[width=0.54\textwidth]{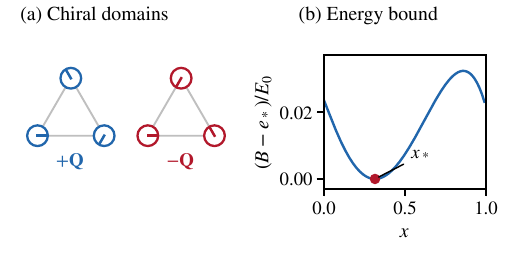}
\caption{Chiral order and its coherent-state energy bound at $g=0.1$. (a) Phases of one dipole component at $\mathbf R$, $\mathbf R+\mathbf a_1$, and $\mathbf R+\mathbf a_2$, ordered counterclockwise. The two patterns are time-reversed domains. (b) The bound $B(x)$ of Eq.~\eqref{eq:bound} reaches the ordered-state energy $e_*$ at $x_*=0.315$. For $x\le0.99$, this auxiliary function bounds the energy of arbitrary coherent configurations after minimizing over the outer occupations. The endpoint region $x>0.99$ is bounded separately in the text. Energies are per cell per reference particle.}
\label{fig:orderbound}
\end{figure}

\section{Quadratic dipole theory and band topology}
\subsection{Construction and conventions}
In the rotating frame of Eq.~\eqref{eq:z}, let $z=(\sqrt{x},\sqrt{y},\sqrt{y},\sqrt{y})^T$ and choose a real $4\times3$ tangent matrix $V$ with $V^TV=I_3$ and $V^Tz=0$. A canonical parametrization is
\begin{equation}
 z_{\mathbf R}^{\rm rot}=z\sqrt{1-\xi_{\mathbf R}^\dagger\xi_{\mathbf R}}+V\xi_{\mathbf R},
 \qquad \xi_{\mathbf R}=a_{\mathbf R}/\sqrt N.
\end{equation}
Expanding the fully specified Eq.~\eqref{eq:Ecl} to second order gives
\begin{equation}
 H_2=\frac12\sum_{\mathbf q}\begin{pmatrix}a_{\mathbf q}^\dagger&a_{-\mathbf q}^T\end{pmatrix}
 \HH(\mathbf q)\begin{pmatrix}a_{\mathbf q}\\a_{-\mathbf q}^\dagger\end{pmatrix}+\mathrm{const},\qquad
 \HH(\mathbf q)=\begin{pmatrix}A(\mathbf q)&B(\mathbf q)\\B(\mathbf q)^\dagger&A(-\mathbf q)^T\end{pmatrix}.
 \label{eq:BdG}
\end{equation}
Equivalently, $A_{ab}(\mathbf q)=\sum_{\mathbf R}e^{2\pi i\mathbf q\cdot\mathbf R}\partial^2E_{\rm cl}/(\partial\xi_{0a}^*\partial\xi_{\mathbf Rb})$ and $B_{ab}(\mathbf q)=\sum_{\mathbf R}e^{2\pi i\mathbf q\cdot\mathbf R}\partial^2E_{\rm cl}/(\partial\xi_{0a}^*\partial\xi_{\mathbf Rb}^*)$, evaluated at $\xi=0$. These derivatives include the normalization term and the stationary-density shift caused by $g$.

For explicit treatment of the charge correction, set $D_{\alpha a}=z_\alpha V_{\alpha a}$. Its density second derivative $R(\mathbf q)$ has nonzero entries
\begin{align}
 R_{00}&=-12gy^2,\qquad R_{\nu\nu}=-4gx^2\cos(2\pi\mathbf q\cdot\mathbf e_\nu),\nonumber\\
 R_{\nu0}&=-4gxy[1+e^{2\pi i\mathbf q\cdot\mathbf e_\nu}],\qquad R_{0\nu}=R_{\nu0}^*,
\end{align}
which add $D^\dagger RD$ to $A$ and $D^\dagger RD^*$ to $B$. Its first derivative is already included when solving $e_g'(x)=0$. Direct finite differences of the real-space energy independently reproduce the quadratic form; gradient and Hessian tests agree below $10^{-7}$.

The bosonic metric is $\Sig=\operatorname{diag}(I_3,-I_3)$. Positive modes satisfy $\HH w_n=\omega_n\Sig w_n$ and $w_n^\dagger\Sig w_m=\delta_{nm}$; the symplectic inner product is $\langle u,v\rangle_{\Sig}=u^\dagger\Sig v$. For $\mathbf q\ne0$, positive-definite $\HH$ permits the Colpa construction~\cite{Colpa1978}: factor $\HH=\mathsf L\mathsf L^\dagger$, and let $c_n$ be a Euclidean unit eigenvector of $\mathsf L^\dagger\Sig\mathsf L$ with positive eigenvalue $\omega_n$. Then $w_n=\sqrt{\omega_n}\,\mathsf L^{-\dagger}c_n$ has unit positive symplectic norm. At $\Gamma$ the Hessian is semidefinite. The two symmetry directions are treated as zero modes, and the isolated upper eigenvector is obtained from $\Sig\HH$ and normalized separately. Thus no positive-definite factorization is applied to the Goldstone null space. The two broken commuting generators give two type-A modes~\cite{WM2012}.

\subsection{Chern invariant, time reversal, and controls}
The band path in Fig.~2 of the Letter is $\Gamma$--$K$--$M$--$\Gamma$, with $\Gamma=(0,0)$, $K=(2/3,1/3)$, and $M=(1/2,1/2)$ in the reciprocal basis. For positive-norm $w_n$, use $\mathcal A_n=i w_n^\dagger\Sig\nabla_{\mathbf q} w_n$ and $C_n=(2\pi)^{-1}\int\Omega_n\,d^2q$~\cite{Shindou2013,Tesfaye2025}. On a periodic mesh, links are normalized overlaps $U_i(\mathbf q)=\langle w(\mathbf q),w(\mathbf q+\hat i)\rangle_{\Sig}/|\langle w(\mathbf q),w(\mathbf q+\hat i)\rangle_{\Sig}|$. The oriented plaquette flux is
\begin{equation}
 F(\mathbf q)=-\arg\left[U_1(\mathbf q)U_2(\mathbf q+\hat1)U_1(\mathbf q+\hat2)^*U_2(\mathbf q)^*\right],
 \qquad C=\frac1{2\pi}\sum F.
\end{equation}
The minus sign implements the stated Berry convention. We use the lattice-gauge discretization of Ref.~\cite{FHS2005}. Band 3 is isolated everywhere; no separate Chern integers are assigned to the two Goldstone bands touching at $\Gamma$.

At $g=0.1$, $60^2$ and shifted $96^2$ meshes give $C_3=-1$ in the $+\mathbf Q$ state. Multi-start momentum minimization gives $\Delta_{23}\simeq0.00606E_0$ near $(0.51430,0.49026)$. Exact time reversal gives
\begin{equation}
 \HH_{-\chi}(-\mathbf q)=\HH_\chi(\mathbf q)^*,\qquad
 \omega_{n,-\chi}(-\mathbf q)=\omega_{n,\chi}(\mathbf q),\qquad C_{3,-\chi}=-C_{3,\chi}.
 \label{eq:TR}
\end{equation}
Time reversal therefore relates the two domains at opposite momenta. As a separate control, $(v,\delta_0,g)=(0,0.2,0)$ with the remaining parameters unchanged gives chiral order but $C_3=0$.

\subsection{Four-node mechanism reference at \texorpdfstring{$g=0$}{g=0}}
Replace the $B$ conversion term with both spatial orientations of its offsets, weighted by $j_B(1+s)/2$ and $j_B(1-s)/2$. The couplings remain real for $|s|\le1$. Both orientations are phase compatible with the chiral saddle, and its energy is independent of $s$. In the real tangent basis, $\HH_s=\Rea\HH_1+is\Ima\HH_1$, with real and imaginary parts taken entrywise. At $s=0$, four linear contacts join bands 2 and 3 (Table~\ref{tab:nodes}).
\begin{table}[htbp]
\centering
\caption{Band-2/3 contacts in the mirrored-conversion reference at $g=0$, $s=0$, and $\chi=+1$. Reduced coordinates lie in $[0,1)$, with $t_0\simeq0.068754$; Fig.~3(b) of the Letter displays equivalent representatives in $[-1/2,1/2)$.}
\label{tab:nodes}
\begin{tabular}{lll}
\toprule
Contact & $\mathbf q$ & $\omega_D/E_0$\\\midrule
Outer & $(1/3,2/3)$ & 1.6169\\
Inner (three) & $(t_0,1-t_0),(1-2t_0,1-t_0),(t_0,2t_0)$ & 0.3651\\
\bottomrule
\end{tabular}
\end{table}
Let $W$ contain the two touching positive modes with $W^\dagger\Sig W=I_2$. The complete first-order projection is
\begin{align}
 h_D={}&[\omega_D+\mathbf v_0\cdot\delta\mathbf q]I_2
 +(\mathbf v_1\delta q_1+\mathbf v_2\delta q_2+s\mathbf m)\cdot\boldsymbol\sigma
 +O(|\delta\mathbf q|^2,s|\delta\mathbf q|,s^2),\nonumber\\
 v_{0i}={}&\tfrac12\operatorname{Tr}[W^\dagger(\partial_{q_i}\HH_0)W].
 \label{eq:diracfull}
\end{align}
The traceless parts of $W^\dagger(\partial_{q_i}\HH_0)W$ and $W^\dagger(\HH_1-\HH_0)W$ define $\mathbf v_i$ and $\mathbf m$. There is no linear scalar term in $s$: at $s=0$, $W$ can be chosen real, whereas $\partial_s\HH_s$ is purely imaginary and antisymmetric. The scalar velocities, in units of $E_0$ for dimensionless $q_i$, are
\begin{equation}
 \mathbf v_0/E_0=(0,0),\quad (\upsilon,-\upsilon),\quad(-\upsilon,0),\quad(0,\upsilon),
 \qquad \upsilon\simeq1.972,
\end{equation}
in the order of the four node coordinates listed above. Central differences with steps from $2\times10^{-5}$ to $2.5\times10^{-6}$ converge to these values. Along $\delta\mathbf q=\epsilon(0.7,-0.4)$, $s=\epsilon$, halving $\epsilon$ gives quadratic frequency errors for the complete expansion. The scalar tilt shifts both branches equally and leaves their eigenvectors and local Berry flux unchanged.

For small positive $s$, the upper-branch flux is $-\tfrac12\operatorname{sgn}[(\mathbf v_1\times\mathbf v_2)\cdot\mathbf m]$. The triple product is $-8.15\times10^{-6}$ at the outer node and $+4.69\times10^{-4}$ at each inner node. Thus the four local contributions sum to $-1$.

At fixed $\chi=+1$ and $g=0$, the interpolation obeys $\HH_{-s}(\mathbf q)=\HH_s(\mathbf q)^*$. Since $\Sig$ is real, complex conjugation preserves the positive frequencies and reverses the Berry curvature. Hence $\Delta_{23}(-s)=\Delta_{23}(s)$ and $C_3(-s)=-C_3(s)$ whenever the third band is isolated. This relation supplies the negative-$s$ points in Fig.~3(c) of the Letter. Independent $60^2$ and shifted $96^2$ meshes at $s=-1$ both give $C_3=+1$, with the same minimum direct gap as at $s=+1$. Reversing $s$ changes the conversion weights while holding the condensate chirality fixed; physical time reversal instead relates opposite chiralities at opposite momenta through Eq.~\eqref{eq:TR}.

\begin{table}[htbp]
\centering
\caption{Reference with mirrored conversion patterns at $g=0$. Gaps are numerical minima obtained by momentum searches. The Chern integer is evaluated over the full Brillouin zone.}
\label{tab:mirrored}
\begin{tabular}{ccccc}\toprule
$s$ & 0.1 & 0.3 & 0.6 & 1\\\midrule
$\Delta_{23}/E_0$ & 0.00163 & 0.00487 & 0.00843 & 0.01024\\
$C_3$ & $-1$ & $-1$ & $-1$ & $-1$\\\bottomrule
\end{tabular}
\end{table}
The Dirac half-integers are local weak-mass contributions, not Chern numbers of separate half-bands. The full lattice invariant and the gap calculations in Table~\ref{tab:mirrored} establish the topology at the sampled nonzero $s$.

\subsection{Continuation to the final charge-projected model}
At fixed physical detuning and $s=1$, we reoptimize the stationary density for each $g=0,0.01,\ldots,0.1$. A $24^2$ reciprocal grid supplies multiple starting points for local gap minimization; additional searches follow earlier minima and the independently known endpoint minimum. The latter resolves a narrow avoided crossing that becomes the smallest gap near the final benchmark. Independent shifted $72^2$ Chern meshes give $C_3=-1$ at every point. The smallest sampled Hessian eigenvalue is positive away from $\Gamma$, where the two symmetry modes remain gapless. Selected values are listed in Table~\ref{tab:continuation}.
\begin{table}[htbp]
\centering
\caption{Continuation from the $g=0$ reference to the $g=0.1E_0$ benchmark at fixed microscopic detuning and $s=1$. The stationary central occupation is reoptimized at each point; all listed points have $C_3=-1$.}
\label{tab:continuation}
\begin{tabular}{ccc}\toprule
$g/E_0$ & $x_*$ & $\Delta_{23}/E_0$\\\midrule
0.00 & 0.3000 & 0.01024\\
0.02 & 0.3031 & 0.01019\\
0.04 & 0.3062 & 0.01013\\
0.06 & 0.3092 & 0.01004\\
0.08 & 0.3121 & 0.00806\\
0.10 & 0.3149 & 0.00606\\
\bottomrule\end{tabular}
\end{table}
This sampled numerical continuation supports a gap-preserving connection of the four-contact reference to the final benchmark.

\section{Physical density spectra and preservation of the Chern number}
\subsection{Smooth probes and site-resolved sources}
For physical wave vector $\mathbf k=\sum_iq_i\mathbf b_i$, define $\rho_{\mathbf k}=\sum_{\mathbf R\alpha}e^{-i\mathbf k\cdot\mathbf r_{\mathbf R\alpha}}n_{\mathbf R\alpha}$. The translation phases are $k_i\equiv\mathbf k\cdot\mathbf a_i=2\pi q_i$, not Cartesian components. Local particle-number and dipole conservation eliminate the constant and linear terms of $[H,\rho_{\mathbf k}]$, giving $O(|\mathbf k|^2)$ transition amplitudes and $O(|\mathbf k|^4)$ weights at finite frequency~\cite{Zechmann2024}. For example, the spatial factor of $T$ is
\begin{equation}
 2e^{-i\mathbf k\cdot\mathbf S}\left[1-\cos(\mathbf k\cdot\mathbf e_\nu/2)\right].
\end{equation}
A uniform force $-\mathbf F(t)\cdot\mathbf P$ cannot induce finite-frequency transitions. Independent site potentials within a cell avoid this restriction. Related probes include density spectra of constrained bosons~\cite{Zechmann2024} and synthetic tensor electric fields~\cite{Zhang2026Resonant}; here we use the physical site densities of Eq.~\eqref{eq:HC}.

Define cell-periodic densities $n_\nu(\mathbf q)=L^{-1/2}\sum_{\mathbf R}e^{-2\pi i\mathbf q\cdot\mathbf R}n_{\mathbf R\nu}$ for the three outer sites. For a resolved band, absorption by $H_f=\epsilon e^{-i\omega t}\sum_\nu f_\nu n_\nu(-\mathbf q)+\hc$ measures, after division by a common drive and exposure factor, the dimensionless weight $W_n(f)=f^\dagger S_n f$ with $S_n=d_nd_n^\dagger$. The transition amplitudes are $(d_n)_\nu=\langle0|n_\nu(\mathbf q)|n,\mathbf q\rangle/\sqrt N$, $\nu=1,2,3$, where $|0\rangle$ is the quasiparticle vacuum. For a weak drive applied for time $t_{\rm obs}$, the frequency-integrated transition probability is $\int d\omega\,P_n(\omega)=2\pi|\epsilon|^2Nt_{\rm obs}W_n(f)$ in the long-time, resolved-band limit. An absorbed-energy measurement includes an additional factor $\omega_n$, which must be removed to obtain the same weight. All complex $f_\nu$ are realized as relative temporal phases of real potentials. These sources commute with $Q$ and $\mathbf P$.

\subsection{Nonvanishing density projection and the Chern number}
Using the real condensate vector $z$ and tangent matrix $V$ defined in Sec.~3.1, a Bogoliubov mode $w_n=(u_n,v_n)^T$ has density amplitude
\begin{equation}
 d_n=M(u_n+v_n),\qquad M_{\nu a}=z_\nu V_{\nu a},\quad \nu=1,2,3.
 \label{eq:densitymap}
\end{equation}
If all $z_\alpha$ are nonzero, $M$ is nonsingular. Indeed, $Mc=0$ forces all outer components of $Vc$ to vanish; its orthogonality to $z$ then forces the central component to vanish, so $c=0$. For the uniform state, $VV^T=I_4-zz^T$ gives $MM^T=y(I_3-y\boldsymbol 1\boldsymbol 1^T)$, where $\boldsymbol 1=(1,1,1)^T$ and $y=(1-x)/3$. Its eigenvalues are $y,y,xy$, so $|\det M|=\sqrt{x}\,y^{3/2}>0$ for $0<x<1$. Positive Krein norm $u_n^\dagger u_n-v_n^\dagger v_n=1$ forbids $u_n+v_n=0$. Thus the density amplitude is everywhere nonzero on any isolated positive band.

The projection-invariance principle and partial tomography have an established precedent~\cite{Huang2018}; smooth topology-preserving maps also underlie the classification of bosonic pairing Hamiltonians~\cite{Chaudhary2021}. Here nonvanishing follows from the physical density map and positive norm. Under a patch gauge change $w_n^{(b)}=e^{i\theta_{ab}}w_n^{(a)}$, the amplitude obeys $d_n^{(b)}=e^{i\theta_{ab}}d_n^{(a)}$, preserving line-bundle transition functions.

\begin{samepage}
With $\partial_i=\partial/\partial q_i$ and $\widehat d_n=d_n/\|d_n\|$, the difference between $i\widehat d_n^\dagger\partial_i\widehat d_n$ and the symplectic connection is a globally defined one-form whose curvature integrates to zero on the Brillouin torus. Thus the measured projector $P_n=S_n/\operatorname{Tr}S_n=\widehat d_n\widehat d_n^\dagger$ gives
\begin{equation}
 C_n^{\rm density}=\frac{i}{2\pi}\int_{\rm BZ}
 \operatorname{Tr}\!\left(P_n[\partial_1P_n,\partial_2P_n]\right)d^2q
 =C_n^{\rm BdG},
 \label{eq:densitychern}
\end{equation}
The density and symplectic local curvatures can differ while their Brillouin-zone integrals agree.
\par\end{samepage}

The same line-bundle argument applies to any smooth density map that respects reciprocal-boundary sewing and is nonzero on the isolated positive-norm Gaussian eigenline. For maps of the form $d=M(u+v)$, full column rank of $M$ and positive Bogoliubov norm guarantee nonvanishing. Full row rank of a generic map from the six-component Bogoliubov space alone does not guarantee it.

The cell-periodic convention is important. The spatially embedded density amplitudes acquire the known diagonal phase $e^{-i\mathbf k\cdot\mathbf e_\nu/2}$. This representation has nontrivial reciprocal-boundary sewing; one must either undo these phases or use the corresponding sewing matrices. Naively identifying opposite edges of the embedded representation would change the numerical problem.

\subsection{Nine-pattern reconstruction and ideal tests}
Let $e_a$ be the three unit drive vectors. Measure
\begin{equation}
 f=e_a,\qquad f_{ab}^{R}=(e_a+e_b)/\sqrt2,\qquad
 f_{ab}^{I}=(e_a+ie_b)/\sqrt2,\qquad a<b.
\end{equation}
The diagonal elements are $S_{aa}=W(e_a)$, and
\begin{align}
 \Rea S_{ab}&=W(f_{ab}^{R})-\tfrac12(S_{aa}+S_{bb}),\nonumber\\
 \Ima S_{ab}&=\tfrac12(S_{aa}+S_{bb})-W(f_{ab}^{I}).
\end{align}
The areas must share a common calibration; individually normalizing each drive spectrum would destroy these relations. Diagonalizing the reconstructed rank-one matrix gives its spectral projector, from which the lattice-gauge Chern calculation uses ordinary overlaps. The inversion uses only the calibrated peak areas and the known drive vectors.

\begin{table}[htbp]
\centering
\caption{Noiseless, band-resolved density tomography. The benchmark has $g=0.1$; the trivial control is defined in Sec.~3.2. The reciprocal mesh is $q_i=(j_i+0.319)/N_q$, with $j_i=0,\ldots,N_q-1$.}
\label{tab:density}
\begin{tabular}{lccc}\toprule
Model & Mesh & $C_3^{\rm density}$ & $C_3^{\rm BdG}$\\\midrule
Benchmark, $+\mathbf Q$ & $48^2$ & $-1$ & $-1$\\
Benchmark, $+\mathbf Q$ & $96^2$ & $-1$ & $-1$\\
Benchmark, $-\mathbf Q$ & $96^2$ & $+1$ & $+1$\\
Chiral trivial control & $48^2$ & $0$ & $0$\\\bottomrule
\end{tabular}
\end{table}
Table~\ref{tab:density} compares the reconstructed and Bogoliubov Chern integers. On the benchmark $96^2$ mesh, $\operatorname{Tr}S_3$ ranges from $0.0537$ to $0.134$. The singular values of $M$ are approximately $0.478$, $0.478$, and $0.268$, consistent with the analytic invertibility. At $\Gamma$ the upper-band density weight is $0.0639$. The control has nonzero local circular contrast despite zero Chern number, separating chirality-sensitive absorption from topology.

At $\mathbf q=(0.839979,0.819146)$, the two circular patterns $(e_1\pm ie_2)/\sqrt2$ give $W_+\simeq0.0556$ and $W_-\simeq0.00287$ at $\omega_3\simeq0.515E_0$. Time reversal gives $S_{3,-\chi}(-\mathbf q)=S_{3,\chi}(\mathbf q)^*$, exchanging these patterns at opposite momentum. Figure 4 of the Letter uses exact line areas. Finite instrumental resolution, counting noise, and dressed high-charge spectral vertices are not included in this ideal reconstruction. The complete microscopic density sum rule also contains high-charge excitations; the effective dipole theory alone does not exhaust this sum rule.

The projector formula uses the rank-one spectral residue of a resolved Gaussian mode. A frequency window containing several modes or a continuum generally yields a higher-rank positive matrix $S_{\rm win}$. Its normalization $S_{\rm win}/\operatorname{Tr}S_{\rm win}$ is not generally a projector, and the projector formula then carries no integer guarantee. A finite-lifetime extension requires a separately justified, continuously isolated nonzero mode-residue line across the Brillouin zone.

\section{Finite occupation, charge corrections, and spectral control}
\label{sec:quantum}
Two parameters control corrections to the quadratic dipole theory: the physical occupation $N$ and the cell-charge gap $N\Lambda$. We treat their effects separately before examining the frequency-resolved spectra.
\subsection{Exact coherent-state factors at finite occupation}
For a fixed-$N$ coherent state, falling-factorial moments obey
\begin{equation}
 \left\langle\prod_\alpha(n_\alpha)_{r_\alpha}\right\rangle=(N)_{\sum r_\alpha}\prod_\alpha p_\alpha^{r_\alpha}.
\end{equation}
Thus the on-site two-body energies have factor $1-1/N$, and the same-cell $A$ conversion term has $(1-1/N)(1-2/N)$. Intercell exchange and the three-cell $B$ conversion term retain their leading factors. For Eq.~\eqref{eq:proj2}, let $A=p_{\mathbf R\nu}$, $B=p_{\mathbf S\nu}$, $C=p_{\mathbf S0}$, and $N_j=(N)_j$. The exact per-particle coherent correction per link is
\begin{equation}
 F_N(A,B,C)=-\frac{g}{N^4}\left[(2NA+1)N_3BC^2+4NAN_2BC+2N^2AB+(NA+1)N_2C^2\right].
\end{equation}
It approaches $-2gABC^2$. The finite-$N$ stationary density is recomputed before taking the Hessian (Table~\ref{tab:finiteN}).
\begin{table}[htbp]
\centering
\caption{Finite-$N$ coherent-state and quadratic theory at second-order charge projection. These entries do not include dynamic self-energy. A dash means the Chern integer was not separately evaluated at that $N$.}
\label{tab:finiteN}
\begin{tabular}{rrrr}\toprule
$N$ & $x_N$ & $\min\Delta_{23}/E_0$ & $C_3$\\\midrule
32 & 0.3762 & 0.00727 & $-1$\\
64 & 0.3467 & 0.00648 & ---\\
128 & 0.3312 & 0.00622 & $-1$\\
256 & 0.3231 & 0.00612 & ---\\
512 & 0.3190 & 0.00609 & ---\\
1024 & 0.3170 & 0.00607 & $-1$\\
$\infty$ & 0.3149 & 0.00606 & $-1$\\\bottomrule
\end{tabular}
\end{table}

\subsection{Physical \texorpdfstring{$1/N$}{1/N} expansion and the one-loop self-energy}
Rotate the physical operators into the condensate and tangent basis, $\widetilde b=(z,V)^\dagger b^{\rm rot}$, and define $E_{\mu\nu}=\widetilde b_\mu^\dagger\widetilde b_\nu$, with $\mu,\nu=0,1,2,3$. For tangent indices $a,b=1,2,3$ and total tangent occupation $n_{\rm ex}=\sum_a a_a^\dagger a_a$, their exact Holstein--Primakoff representation is
\begin{equation}
 E_{00}=N-n_{\rm ex},\quad E_{0a}=\sqrt{N-n_{\rm ex}}\,a_a,\quad
 E_{a0}=a_a^\dagger\sqrt{N-n_{\rm ex}},\quad E_{ab}=a_a^\dagger a_b.
\end{equation}
Normal ordering the microscopic operators, including Eq.~\eqref{eq:proj2}, generates
\begin{equation}
 H=NE_{\rm cl}+E_{\rm const}^{(1)}+H_2
 +N^{-1/2}(H_3+H_{1,\rm ord})+N^{-1}(H_4+H_{2,\rm ord}+E_{\rm const}^{(2)})+\cdots.
\end{equation}
The finite-$N$ coherent factors above provide an independent check; they are not added again to this normal-ordered expansion. Contracting $H_3+H_{1,\rm ord}$ gives a force $F$ and saddle shift $\Xi=-\HH(0)^+(F,F^*)^T$ in the nonzero-stiffness subspace, where $\HH(0)^+$ is the Moore--Penrose pseudoinverse. The shift in $H_3$, quartic contraction, and $H_{2,\rm ord}$ give the static self-energy. The two Goldstone directions satisfy the static--dynamic Ward cancellation, with a numerical matrix residual $2.6\times10^{-16}$ in the stated normalization.

Define fully symmetrized decay and source vertices by
\begin{align}
 H_3^{\rm phys}=\frac1{\sqrt{NL}}\bigg[&\frac12\sum_{\mathbf q\mathbf p,n\mu\nu}
 V_{\mu\nu;n}(\mathbf p,\mathbf q-\mathbf p;\mathbf q)
 \beta_{\mathbf p\mu}^\dagger\beta_{\mathbf q-\mathbf p,\nu}^\dagger\beta_{\mathbf q n}
 \nonumber\\
 &+\frac16\sum_{\mathbf k_1+\mathbf k_2+\mathbf k_3=\mathbf G,n\mu\nu}
 W_{n\mu\nu}\beta_{\mathbf k_1n}^\dagger\beta_{\mathbf k_2\mu}^\dagger\beta_{\mathbf k_3\nu}^\dagger+\hc\bigg].
\end{align}
With Brillouin-zone measure normalized to one, the positive-mode self-energy coefficient is
\begin{align}
 \Sigma_{ab}^R(\mathbf q,\omega)={}&\Sigma_{{\rm st},ab}
 +\frac12\int_{\rm BZ}\sum_{\mu\nu}
 \frac{V_{\mu\nu;a}^*V_{\mu\nu;b}}{\omega-\omega_\mu(\mathbf p)-\omega_\nu(\mathbf q-\mathbf p)+i0}\nonumber\\
 &-\frac12\int_{\rm BZ}\sum_{\mu\nu}
 \frac{W_{a\mu\nu}W_{b\mu\nu}^*}{\omega+\omega_\mu(\mathbf p)+\omega_\nu(-\mathbf q-\mathbf p)}.
 \label{eq:self}
\end{align}
The physical correction is $\Sigma^R/N$. The decay term is the multi-mode Beliaev process~\cite{Beliaev2024}. A Hermitian on-shell approximation uses
$S_{ab}^{\rm os}=\{[\operatorname{Herm}\Sigma(\omega_a)]_{ab}+[\operatorname{Herm}\Sigma(\omega_b)]_{ab}\}/2$.
For a matrix $X$, $\operatorname{Herm}X=(X+X^\dagger)/2$ and $\operatorname{Im}_{\rm H}X=(X-X^\dagger)/(2i)$. The modal projection of $-\operatorname{Im}_{\rm H}\Sigma^R/N$ estimates a half width only when a narrow-peak approximation is valid.

The internal two-dimensional integral uses triangles with linearly interpolated energies and matrix numerators. Principal values and energy-conserving delta weights are integrated analytically, without an artificial linewidth. At the narrowest crossing, internal grids $48^2,96^2,192^2,384^2$ were compared. The last change in real and damping matrix coefficients is below $0.0023$, or $2.3\times10^{-7}E_0$ after division by $N=10^4$. Other crossings were checked with $96^2$ and $192^2$ grids. These comparisons estimate numerical convergence at the sampled momenta.

\subsection{Third- and fourth-order charge processes}
Three forward transfers around an oriented cell triangle return to the subspace with $N_{\mathbf R}=N$ in every cell. Six time orderings have the same pair-defect denominator, giving
\begin{equation}
 H_{\rm eff}^{(3)}=-\frac{6t^3}{N^5\Lambda^2}\sum_{\triangle}(\prod_{e\in\triangle}T_e+\hc),
 \qquad e_3=-24\frac{t^3}{\Lambda^2}(xy)^3.
\end{equation}
At fixed $g=0.1$, reoptimization with this correction yields minimum gaps $0.00514$, $0.00577$, and $0.00596$ in units of $E_0$ for $\Lambda/E_0=10,100,1000$. Thus a finite charge gap reduces the narrow gap.

For $\Delta=\Lambda/2$ and $w=\Lambda/4$, the connected fourth-order classical energy, with $C_e=z_{\mathbf S0}^{*2}z_{\mathbf R\nu}z_{\mathbf S\nu}$, is
\begin{equation}
 E_4=\frac{t^4}{\Lambda^3}\left[\frac{67}{17}\sum_e|C_e|^4
 -\frac49\sum_{\substack{e,f\ \mathrm{adjacent}\\\mathrm{unordered}}}|C_e|^2|C_f|^2
 -40\sum_{\lozenge}\Rea\prod_{e\in\lozenge}C_e^{\rm oriented}\right].
\end{equation}
Disconnected paths cancel the normalization terms. Independent enumeration on a periodic $6\times6$ cell lattice with random complex links reproduces this expression within $5\times10^{-13}$. On the uniform zero-flux background,
$e_4=-(5857/51)(t^4/\Lambda^3)(xy)^4$.
At $\Lambda=1000E_0$ it changes the tracked narrowest gap by $-6.55\times10^{-7}E_0$. The first energy-denominator insertion from $H_n$ has a conservative coherent Hessian bound $1.09\times10^{-7}E_0$ at $N=10^4$. This bound covers that insertion, not the sum of all omitted orders.

\subsection{Frequency-resolved peaks and the limit of uniform spectral control}
For $N=10^4$, $\Lambda=1000E_0$, $t=10E_0$, and the cell-occupation parameters above, the corrected narrowest gap is about $0.00595E_0$; the sum of on-shell band-2/3 half widths there is $0.00448$ of the gap. Other momentum regions have appreciably larger damping. We therefore also examine $N=10^5$ and retain frequency dependence:
\begin{equation}
 G^R(\mathbf q,\omega)=\left[\omega I-\operatorname{diag}\omega^{(4)}(\mathbf q)-\Sigma^R(\mathbf q,\omega)/N\right]^{-1},
 \quad\mathcal A=-\frac{G^R-G^{R\dagger}}{2\pi i}.
 \label{eq:GF}
\end{equation}
Here $\omega^{(4)}$ includes the fourth-order correction from virtual charge fluctuations, and the vertices include third-order charge processes. This Dyson resummation retains the positive-sector normal one-loop self-energy. The plotted scalar spectrum is $A_n(\omega)=c_n^\dagger\mathcal A(\omega)c_n$, where $c_n$ is a Euclidean unit eigenvector of the Hermitian on-shell matrix $\operatorname{diag}\omega^{(4)}+S^{\rm os}/N$. The on-shell frequencies and vertices entering $S^{\rm os}$ include charge processes through third order. This projection labels the corrected modes within the three-dimensional positive-mode sector; it does not supply a full interacting Nambu Green function.

On a $384^2$ internal grid at $N=10^5$, the spectra at $\mathbf q_0=(0.280917,0.447583)$ have band-2 and band-3 peaks near $1.30625E_0$ and $1.31950E_0$. Their respective half widths are $8.69\times10^{-5}E_0$ and $3.55\times10^{-5}E_0$. Their sum is $0.924\%$ of the separation; a shifted mesh gives $0.926\%$. At $N=10^4$ the corresponding spectra are asymmetric and the ratio is about $11.6\%$. Two other checked broad-spectrum points at $N=10^5$ give approximately $0.80\%$ and $0.94\%$.

For each band, the peak frequency is the highest maximum within its calculated frequency interval. The half width is $(\omega_R-\omega_L)/2$, where $\omega_L$ and $\omega_R$ are the nearest half-height crossings on either side of that maximum. The width ratio below is the sum of the two half widths divided by the separation of the dominant peaks. If a secondary peak remains connected above half height, the width encloses both peaks; it is then a spectral-width estimator rather than a single-pole lifetime.

\begin{figure}[htbp]
\centering\includegraphics[width=\textwidth]{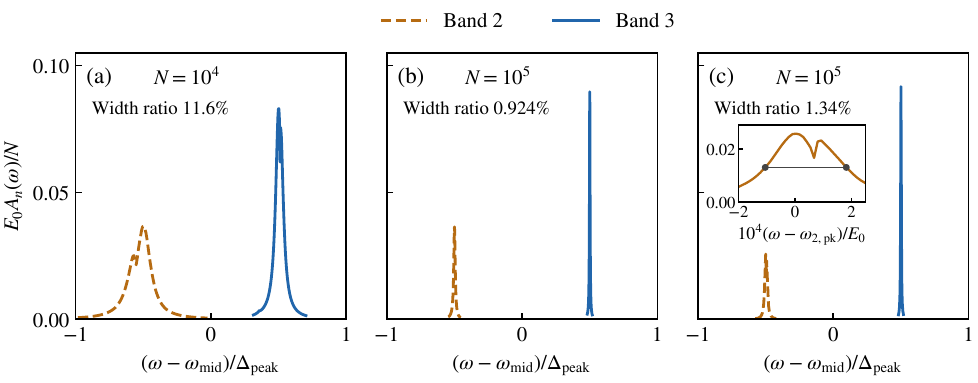}
\caption{Frequency-resolved one-loop spectra from Eq.~\eqref{eq:GF}, with no artificial linewidth. (a,b) The broad-spectrum point $\mathbf q_0$ at two physical occupations. (c) The band-2 saddle resonance at $\mathbf q=(0.282822,0.447583)$ and $N=10^5$. Here $\omega_{\rm mid}$ is the mean of the two peak frequencies and $\Delta_{\rm peak}$ their separation. The common amplitude divisor $N$ makes the two occupations comparable; the two bands are not independently normalized. Curves are drawn only within the calculated frequency intervals. The inset in (c) resolves the band-2 secondary peak; its horizontal segment and endpoint dots mark the half-height interval. Here $\omega_{2,{\rm pk}}$ is the dominant band-2 peak frequency. Width ratios use the half-height estimator defined in the text. All panels use a $384^2$ internal mesh.}
\label{fig:spectra}
\end{figure}

\begin{samepage}
A two-particle saddle places a stronger restriction on the on-shell approximation. If its local energy is $E_s+(h_+X^2-h_-Y^2)/2$, the density of states has $\rho(E)\sim2\log(E_c/|E-E_s|)/\sqrt{h_+h_-}$. Nonzero vertices then produce a logarithmically singular bare on-shell coefficient when a mode crosses $E_s$. Such resonances occur near the points above. Frequency-resolved spectra at $N=10^5$ still display separated peak groups at the tested resonance points, with half-width-sum/separation ratios of about $1.03\%$ and $1.34\%$ on $384^2$ grids; the band-2 resonance has a secondary peak.
\par\end{samepage}

Hermitian one-loop continuations on $36^2$ and $60^2$ external meshes retain $C_3=-1$. These calculations support the leading collective topology and finite-parameter spectral separation at the checked momenta. They do not establish a uniformly narrow, continuously identifiable interacting band over the entire resonance curve, or an interacting Green-function topological invariant.

\section{Conserving wave-packet response}
\label{sec:energy}
A conserving spatial modulation can change collective propagation while the total particle dipole remains fixed. We calculate the resulting transverse displacement of a specified quadratic excitation-energy center; Sec.~6.4 identifies the additional terms needed to relate it to a microscopic energy-density measurement.
\subsection{Drive, modes, and geometric force}
For a closed system the exact particle dipole obeys $\dot P_i=\int j_i\,d^2r=0$. Local higher-rank charge continuity and ordinary energy continuity are compatible~\cite{Gromov2020,Burchards2022,Guo2022}:
\begin{equation}
 \partial_t n+\partial_i\partial_jJ_{ij}=0,\qquad
 \partial_t e+\partial_i j_i^E=0,\qquad E\dot{\mathbf R}_E=\int\mathbf j^E\,d^2r.
\end{equation}
The last relation assumes negligible boundary flux and a specified local energy assignment. Conservation of the particle center does not constrain the energy center to remain fixed.

Use the physical density modulation
\begin{equation}
 H_\eta=\sum_{\mathbf R}\frac{\eta_{\mathbf R}}{2N}(n_{\mathbf R0}-Nx_*)^2,
 \qquad\eta_{\mathbf R}=G f(R_1),\quad f(R_1)=\frac{L_v}{2\pi}\sin\frac{2\pi R_1}{L_v}.
\end{equation}
The leading classical symbol has zero first derivative at $x_*$, so the inhomogeneity preserves the leading saddle and changes its fluctuation stiffness. In the Gaussian theory,
\begin{equation}
 i\partial_tW=\Sig\HH W,\qquad \HH=\HH_0+\eta_{\mathbf R}K,\qquad
 K=dd^\dagger,\quad d=(D_0^*,D_0)^T,\quad D_{0a}=z_0V_{0a}.
\end{equation}
The complete mode basis satisfies $\HH w_m=\epsilon_m\Sig w_m$, $w_m^\dagger\Sig w_l=s_m\delta_{ml}$ and $\sum_ms_mw_mw_m^\dagger\Sig=I$, where $s_m=\pm1$ and all six signed frequencies are retained. This section uses translation phases $k_i\equiv\mathbf k\cdot\mathbf a_i=2\pi q_i$ and their conjugate lattice coordinates $R_i$, so $\partial_{k_i}=(2\pi)^{-1}\partial_{q_i}$. Hellmann--Feynman differentiation gives
\begin{equation}
 \zeta_n=\partial_\eta\epsilon_n=w_n^\dagger K w_n=|d^\dagger w_n|^2,
 \qquad\dot k_1=-\zeta_n\partial_{R_1}\eta.
\end{equation}
This is a force on collective energy generated by a conserving interaction.

For $\alpha\in\{k_1,k_2,\eta\}$,
\begin{equation}
 D_\alpha w_n=\sum_{m\ne n}s_mw_m\frac{w_m^\dagger(\partial_\alpha\HH)w_n}{\epsilon_n-\epsilon_m},\qquad
 \Omega_{\alpha\beta}=-2\Ima(D_\alpha w_n)^\dagger\Sig D_\beta w_n.
\end{equation}
The local gradient also changes the packet energy,
\begin{equation}
 \widetilde\epsilon_n=\epsilon_n+g_1\mathcal E_1,\qquad
 \mathcal E_1=-\Ima[w_n^\dagger(K-\zeta_n\Sig)\partial_{k_1}w_n],\quad g_1=\partial_{R_1}\eta.
\end{equation}
The subtraction makes $\mathcal E_1$ gauge invariant. The adiabatic wave-packet action~\cite{Sundaram1999,Zhang2006,Xiao2010,Tesfaye2025} is
$L=\mathbf k\cdot\dot{\mathbf R}_{\Sig}+\mathcal A_{k_i}\dot k_i+\mathcal A_\eta\dot\eta-\widetilde\epsilon_n$,
where $\mathbf R_{\Sig}$ is the Krein-norm center. In triangular coordinates $Y/a=R_2+R_1/2$ is transverse to the $R_1$ gradient. With $\partial_\perp=\partial_{k_2}+\tfrac12\partial_{k_1}$ and $v_1=\partial_{k_1}\epsilon_n$, the geometric correction is
\begin{equation}
 \frac{\delta\dot Y_{\Sig}}{a}=g_1\left[\zeta_n\Omega_{21}+\partial_\perp\mathcal E_1
 -v_1(\Omega_{2\eta}+\tfrac12\Omega_{1\eta})\right].
 \label{eq:velocity}
\end{equation}
The three terms describe momentum Berry deflection, a gradient-energy correction, and changing local mode polarization. A chosen quadratic excitation-energy center also has observable and normalization corrections; it is evaluated below.

\subsection{Quadratic excitation-energy center and finite-time linear response}
Let $h^{\rm eff}_{\mathbf R}$ denote the origin-assigned local density of the charge-projected Hamiltonian, including its second-order density interaction. We define a quadratic excitation-energy convention by retaining the degree-two term $h^{\rm eff}_{\mathbf R,2}$. Its spatial sum is $H_2$. For a monomial with fluctuation offsets $r,s$ relative to that origin, let $M_E$ be its Fourier kernel multiplied by $(r_\perp+s_\perp)/2$, summed over the terms. Then with $X=i(\partial_{k_2}+\tfrac12\partial_{k_1})$,
\begin{equation}
 \mathcal A(G)=\tfrac12\{\HH_0+GV,X\}-M_E,\qquad V=f(R_1)K,\qquad
 R_E^{(2)}(G)=\frac{\langle W_G|\mathcal A(G)|W_G\rangle}{\langle W_G|\HH_0+GV|W_G\rangle}.
 \label{eq:center}
\end{equation}
$M_E$ converts symmetric endpoint assignment to the specified quadratic energy density; the on-site drive has no additional correction. Inner products here are ordinary Nambu inner products, integrated over position or momentum. Equation~\eqref{eq:center} defines the quadratic excitation-energy center above the unexcited Gaussian reference. Its relation to a microscopic energy-density measurement is discussed below.

Expand $W_G=W^{(0)}+GW^{(1)}+O(G^2)$ for a $G$-independent initial packet. Let $E^{(0)}=\langle W^{(0)}|\HH_0|W^{(0)}\rangle$ and $R^{(0)}=\langle W^{(0)}|\mathcal A(0)|W^{(0)}\rangle/E^{(0)}$. Differentiating the observable gives
\begin{align}
 \left.\partial_GR_E^{(2)}\right|_0=\frac1{E^{(0)}}\bigg\{&2\Rea\langle W^{(1)}|[\mathcal A(0)-R^{(0)}\HH_0]|W^{(0)}\rangle\nonumber\\
 &+\langle W^{(0)}|[\tfrac12\{V,X\}-R^{(0)}V]|W^{(0)}\rangle\bigg\}.
 \label{eq:energyresponse}
\end{align}
The second line is the explicit readout term, including the energy denominator. Subtract the derivative at $t=0$ separately for each protocol to form displacements.

No driven propagation is required for the first-order state. For initial amplitude $a(\mathbf q')w_3(\mathbf q')$, the exact finite-time Dyson integral in the Gaussian theory is
\begin{align}
 W^{(1)}(\mathbf q,t)&=\sum_{m,\mathbf q'}s_mw_m(\mathbf q)
 [w_m^\dagger(\mathbf q)V_{\mathbf q\mathbf q'}w_3(\mathbf q')]a(\mathbf q')e^{-i\epsilon_m(\mathbf q)t}
 F_t[\epsilon_3(\mathbf q')-\epsilon_m(\mathbf q)],\nonumber\\
 F_t(\epsilon)&=\frac{e^{-i\epsilon t}-1}{\epsilon}
 =-it e^{-i\epsilon t/2}\operatorname{sinc}(\epsilon t/2),\qquad F_t(0)=-it.
 \label{eq:Dyson}
\end{align}
Only $q'_1=q_1\pm L_v^{-1}$ occur for the sinusoidal modulation. Equations~\eqref{eq:energyresponse}--\eqref{eq:Dyson} include all six modes, the source, switching, and the actual energy assignment.

\subsection{Numerical protocol and quantitative comparison}
Set $G=5\times10^{-5}E_0$, $L_v=1024$, and prepare band 3 at $\mathbf q_0=(0.510,0.480)$ in the $+\mathbf Q$ domain with Gaussian probability width $\sigma_{q_1}=0.002$. The other direction is taken in the narrow-width limit. Use $-\mathbf q_0$ in the time-reversed domain. The local frequency and separation are $\omega_3\simeq1.024E_0$ and $\omega_3-\omega_2\simeq0.0334E_0$. Define
\begin{equation}
 \Delta Y_{\rm odd}(t)=\frac14\sum_{\chi,\tau=\pm1}\chi\tau
 [Y_{\chi,\tau G}(t)-Y_{\chi,\tau G}(0)].
\end{equation}
Matched initial distributions and $\eta_{-\tau}(-R_1)=\eta_\tau(R_1)$ remove the ordinary local-dispersion ray contribution. The propagation and mode-based theory use the same local energy, with the response calculation independently evaluated from undriven modes.

\begin{table}[htbp]
\centering
\caption{Quadratic excitation-energy displacement at $t=400\hbar/E_0$. The extrapolation cancels the leading cubic-in-$G$ odd term; it is a finite-gradient estimate of linear response.}
\label{tab:energy}
\begin{tabular}{lr}\toprule
Calculation & $\Delta Y_{\rm odd}/a$\\\midrule
Direct six-mode propagation at $G$ & 0.013906\\
Twice the propagation at $G/2$ & 0.013719\\
Extrapolation $[8Y(G/2)-Y(G)]/3$ & 0.0136572\\
Finite-time six-mode theory & 0.0136538\\\midrule
Original-band state contribution & 0.026162\\
Band-2 coherent contribution & $-0.013546$\\
Explicit energy-readout contribution & 0.000816\\
Other modes & 0.000222\\\bottomrule
\end{tabular}
\end{table}
Table~\ref{tab:energy} shows agreement between the theory and extrapolated propagation within $0.025\%$. The frozen-band result is the original-band term plus explicit readout, approximately $0.02698a$; band-2 coherence is first order in the gradient even when transferred population is small. It cannot be judged by population leakage alone. An adiabatic integration of Eq.~\eqref{eq:velocity} gives $0.02868a$ from momentum curvature, $-0.00278a$ from gradient energy, and $-0.01236a$ from mixed curvature. Their total, $0.01354a$, explains the full result but is not an exact decomposition of the finite-time table. Energy polarization, preparation, and finite-time coherence remain in Eqs.~\eqref{eq:energyresponse}--\eqref{eq:Dyson}. As in other neutral wave-packet responses~\cite{Matsumoto2011,Price2012}, the local displacement is not a measurement of the band Chern integer or a quantized thermal Hall coefficient.

\begin{figure}[!htb]
 \centering\includegraphics[width=0.54\textwidth]{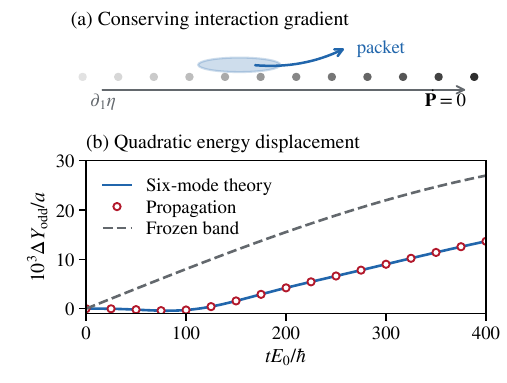}
 \caption{Transverse wave-packet response in the quadratic theory. (a) Schematic of the two-dimensional response: the gray row indicates the interaction-gradient direction, and the curved arrow illustrates transverse packet deflection. The conserving central-site interaction leaves the microscopic particle center fixed, $\dot{\mathbf P}=0$. (b) Chirality- and gradient-odd quadratic excitation-energy displacement: full six-mode linear response, direct propagation extrapolated to linear order, and the frozen-band approximation. The observable uses the origin assignment defined in Eq.~\eqref{eq:center}. Here $G=5\times10^{-5}E_0$, $L_v=1024$, and $\mathbf q_0=(0.510,0.480)$, with opposite momentum in the time-reversed domain.}
 \label{fig:energy}
\end{figure}

\clearpage
\subsection{Relation to a microscopic local-energy measurement}
The quadratic energy convention above is well defined, but projection of the Hamiltonian alone does not determine the projection of a microscopic local energy density.

For a transfer link from $\mathbf R$ to $\mathbf S$, write $A=\mathcal P_0V_{\rm link}\mathcal Q_0(N\Lambda)^{-1}\mathcal Q_0V_{\rm link}\mathcal P_0$, where $V_{\rm link}$ is the microscopic transfer term on this link and $\mathcal P_0$ projects onto $N_{\mathbf R}=N$ in every cell and $\mathcal Q_0=1-\mathcal P_0$. Its second-order effective Hamiltonian is $-A$. Under an origin assignment of the microscopic transfer term, each endpoint cell-occupation energy acquires $+A/2$, whereas the transfer energy at $\mathbf R$ acquires $-2A$. The projected microscopic local contributions are therefore $-3A/2$ at $\mathbf R$ and $+A/2$ at $\mathbf S$. Relative to assigning $-A$ wholly to $\mathbf R$, their energy first moments differ by $(\mathbf S-\mathbf R)A/2$. This polarization term is of order $g$, even at fixed $g$ and asymptotically large charge-gap scale.

A second contribution comes from the excitation-induced mean field. In the local expansion $h_{\mathbf R}=Ne_{\mathbf R}+\sqrt N h_{\mathbf R,1}+h_{\mathbf R,2}+\cdots$, stationarity removes $\sum_{\mathbf R}h_{\mathbf R,1}$ but need not remove its spatial moment. An induced tangent expectation of order $N^{-1/2}$ then contributes to the energy first moment at the same order as $h_{\mathbf R,2}$. Writing the creation part of $h^{\rm eff}_{\mathbf R,1}$ as $\sum_{r,a}c_{ra}a^\dagger_{\mathbf R+r,a}$, define $c_a^\perp=\sum_r(r_2+r_1/2)c_{ra}$, with $Y/a=R_2+R_1/2$. In the projected benchmark, $\|c^\perp\|=0.0341E_0$, although $\sum_r c_{ra}=0$. Phase averaging does not generally eliminate this second-order mean response. These two terms are not included in Fig.~\ref{fig:energy}. Consequently the quoted displacement tests the specified quadratic excitation-energy response, not the complete microscopic energy-center response. Neither correction affects the Gaussian band invariant or the leading density-projection result.

\subsection{Particle dipole and the condensate response}
The microscopic dipole differs from the Krein-norm packet position. Expand it consistently with the Hamiltonian:
\begin{equation}
 H=NE_{\rm cl}+H_2+N^{-1/2}H_3+\cdots,\qquad
 P=NP_0+\sqrt N P_1+P_2+N^{-1/2}P_3+\cdots.
\end{equation}
Exact conservation yields
\begin{equation}
 [H_2,P_1]=0,\qquad [H_2,P_2]+[H_3,P_1]=0.
\end{equation}
The drive is included in this expansion. Thus $[H_2,P_2]$ alone need not vanish: the induced background response supplies $\sqrt N\Delta\langle P_1\rangle+\Delta\langle P_2\rangle=0$ at the retained order. This is the constraint on total particle redistribution. The spatial profile of that compensation is not extracted from the Gaussian packet calculation. Translation of a neutral dipole itself preserves $\mathbf P$, and the conservation law places no prohibition on neutral energy propagation. The computed Gaussian response illustrates collective motion of the dipole condensate; it does not determine the full spatial energy or particle redistribution of an excitation and its induced background.

\small
\bibliographystyle{apsrev4-2}
\bibliography{references}